\documentstyle[preprint,eqsecnum,aps]{revtex}

\begin{document}
\draft
\title{Theory of Coexisting Transverse Spin Freezing \\ and Long-Ranged
       Antiferromagnetic Order \\ in Lightly Doped La$_{2-x}$Sr$_x$CuO$_4$}
\author{R.J. Gooding, N.M. Salem}
\address{Dept. of Physics, Queen's University,\\
Kingston, Ontario, Canada K7L 3N6}
\author{A. Mailhot}
\address{Dep. de Physique, Univ. de Sherbrooke, \\
Sherbrooke, Quebec, Canada J1K 2R1}
\date{\today}
\maketitle
\bigskip
\bigskip
\bigskip
\begin{abstract}
We provide an explanation of the spin--freezing transition recently
observed by Chou $et~al.$ (Phys. Rev. Lett. {\bf 71}, 2323 (1993))
in $La_{2-x}Sr_xCuO_4$ for $x~{< \atop \sim}~.02$. We propose that topological
excitations of the 2D Heisenberg quantum antiferromagnet having non--coplanar
transverse
components have a pair--interaction energy that qualitatively and
quantitatively
agrees with the observed values of spin--freezing temperature as a function
of doping.
\end{abstract}
\pacs{}

\newpage
\section{Introduction:}
\label{sec:intro}
One of the avenues that may aid in the understanding of the properties of
high--temperature superconductors involves examining the magnetic correlations
that exist in the $CuO_{2}$ planes when doped with carriers. Experiments
may be employed to investigate the local magnetic character of these systems -
some of the most
useful are nuclear quadrupole resonance (NQR) studies where the nucleus
provides a local probe\cite{nishihara,borsa}, and from this information many
features of the
magnetic ordering and/or fluctuations may be identified.
Further, through resonance frequencies,
one can measure the coupling between the nucleus and the electric field
gradient surrounding it, thus providing insight into the electronic charge
distribution
of the system.

The parent compound of the Bednorz--M\"uller high--$T_c$ compound,
viz. $La_2CuO_4$, is an antiferromagnetic (AFM) insulator which undergoes
a N\'eel--ordering transition at \hfill\break
$\sim$ 300 K. When the compound has trivalent
$La$ substituted for with divalent $Sr$ for low doping levels, viz.
less than 1\% $Sr$, the AFM ordering persists with the N\'eel transition
temperature ($T_N$) rapidly decreasing with increasing $Sr$ concentration.
Recent $^{139}La$ NQR experiments \cite{chou} on lightly doped
$La_{2-x}Sr_xCuO_4$
($x \leq 0.02$) have suggested that an interesting {\it addition} to the AFM
structure takes place. To be specific, they have measured data consistent with
the coexistence of the AFM phase with long--ranged order
{\it and} a spin--glass phase, and have noted that this latter state can be
interpreted to
be composed of frozen transverse (i.e., perpendicular to the direction of
the staggered magnetization) spin components. The state may be unambiguously
assigned to be that of a spin glass via the stretched exponential
recovery of the nuclear magnetization \cite{chen}. A schematic diagram of the
low $x$ phase diagram is shown in Fig. 1, where a line separating the
pure N\'eel state and the spin--glass coexisting state is denoted by $T_f$; we
shall
refer to this line as the spin--freezing temperature. The focus of this paper
is to (i) suggest the spin texture of this coexisting state, and (ii) show
that this spin texture predicts an onset of coexistence at $T_f$ consistent
with experiment.

To begin, let us summarize the experimental results.
Chou {\it et~al.} \cite{chou} have measured the $^{139}La$ NQR for different
dopings and temperatures in lightly doped $La_{2-x}Sr_xCuO_4$ for low
temperatures.
They measured the $^{139}La$ NQR frequency spectrum, and the relaxation times
for the system's magnetization to return to equilibrium.
The splitting of the $2 \nu_Q$ NQR peaks due to
the internal field associated with
the AFM ordering of the spins at the N\'eel temperature led to a
doping--dependent
ordering temperature $T_N (x)$ - this determination of $T_N (x)$ is in
agreement
with other measurements, such as (i) their observations of the spin--lattice
relaxation rate, and (ii)
that found using static susceptibility measurements \cite{cho}. Further, a
large enhancement
of this splitting was found at very low $T$, viz. at temperatures less than
$\sim$ 30 K,
with the enhancement increasing with increasing $x$.
Such behaviour could be caused by an alteration of the underlying spin texture
at
this temperature.  A change in
the spin state may also be inferred from measurements of the relaxation rate of
the $3 \nu_Q$
transition -  below $\sim 70 K$ there is a large enhancement of this rate.
The rate continues to increase until it rapidly drops off at very low
temperatures, viz. temperatures
around 5 - 16 K.
Using this low temperature behaviour of the relaxation rate, Chou {\it et~al.}
\cite
{chou}
identified the maximum of the NQR rate with the spin--freezing temperature
$T_f$. As a function of doping
it was determined to be
approximately given by $T_f(x) \sim (815 K) x$, and it is this linear relation
that we
have sketched in Fig. 1. Above 30K (which may be interpreted to be the
charge localization transition temperature \cite {chouthesis}) the nuclear
magnetization relaxed with a
multiexponential decay, while below this temperature a dramatic change was
noted in that now
the relaxation of the magnetization was found to be described
by $M(\infty) - M(t) \propto \exp[-(t/T_1^\ast)^{1/2}]$.
This stretched exponential decay has been found in many systems studied
by NQR (see, e.g. Ref. \cite{chen}), and can be associated with a spin--glass
ordering.
However, the staggered moment of the AFM phase does not disappear, as
indicated, e.g., in
(i) the zero--temperature extrapolation of the $2\nu_Q$ splitting to the
undoped value,
or (ii) the continuous change of the splitting as $T_f$
is traversed (see Fig. 1(c) of \cite {chou} for the data corresponding to both
of these results),
and thus the spin glass ordering is found to coexist with the long--ranged AFM
order.

Similar experimental results were found in both other NQR work \cite{watanabe}
at
$x = .012$ and .015, as well as in electron paramagnetic resonance (EPR)
studies of the same system at $x = .009$ by Rettori {\it et~al.} \cite{rettori}
(see their
Fig. 12) - his latter study again found an increase in internal field
observed in the $-7/2~\rightarrow~-5/2$
transition.

Clearly, it would be desirable to explain the above--mentioned behaviour of the
spin texture,
and associated experimental measurements, and that is the focus of this paper.
However,
we wish to stress that this phenomenon is certainly not limited to the weakly
doped Bednorz--M\"uller high $T_c$ compound, nor to AFMs. For example, previous
experimental work
on metallic spin glasses \cite{ryan}, and related computational studies
\cite{thomson}
on the moderately frustrated three-dimensional Heisenberg model, aids in the
understanding of the
high $T_c$ system we are studying.
The work on the moderately frustrated Heisenberg model leads to
the phase diagram shown in Fig. 2 of Ref. \cite{thomson}, and for dopant levels
less than 25 \%
is similar to that of our Fig. 1. In particular, the linear
dependence of $T_{xy}$ is the same. The authors of Ref. \cite{thomson}
identified the coexisting freezing of the transverse spin degrees of freedom
(transverse in the sense of being perpendicular to the already present
ferromagnetic
moment) and long--ranged ferromagnetic order at the spin--freezing temperature
$T_{xy}$;
this latter ordering is not associated with a phase transition, only involving
a short--ranged
freezing of spins, and does not eliminate the ferromagnetic ordering.
Unfortunately,
these authors did not attempt to come to an analytic understanding of this
similar
 linear dependence of
the spin--freezing temperature on doping. Nor did they attempt to justify the
magnitude
of the slope of this pseudo--phase boundary. That is precisely what we will
present
in this paper, albeit for at a doped two--dimensional AFM.

Our paper is organized as follows. In \S II we summarize our previous work
\cite{1skyrmion,nskyrmions} on the spin texture of singly and weakly doped
$CuO_2$ planes; this involves studies of a non--coplanar spin texture where
each $Sr^{2+}$
dopant produces a spin state reminiscent of a singly charged skyrmion.
This section also serves to introduce the effective Hamiltonian
that we will use to describe these doped planes. In \S III we present
an analytic calculation of the two skyrmion interaction energy, and
then use this to suggest the dependence of $T_f(x)$
on doping $x$. In \S IV we describe numerical work
examining the same two skyrmion interaction energy problem, and which
reaches the same qualitative conclusion as in \S III,
viz. an interaction energy $\propto x$, as well as providing a quantitative
estimate of the slope.
Then, we conclude this section by extrapolating the above
work to determine the dependence of the spin--freezing temperature on
doping via both a comparison to Chou $et~al.$ \cite {chou}, and
a simple counting of the degrees of freedom of the two defect
system. Finally, in \S V
we summarize how this and other work on the weakly doped planes
leads to a clear description of the novel spin texture of this system.

\section{Spin Texture of Weakly Doped Planes at Low $T$:}
\label{sect:texture}

Our model of the doped state of a $CuO_2$ plane at low $T$ is based on a
collection
of partially delocalised holes. It is assumed that at low temperatures
the carriers (viz. $O$ holes) are not mobile, but rather are confined
to some small region of a plane - this is entirely consistent with
the localisation transition seen in resistivity measurements between
50 and 100 K \cite{resistivity}, as well as with the localisation
transition inferred from the NQR measurements of Chou \cite {chouthesis}.
Physically, this localisation can be considered to arise from
the electrostatic attraction of the holes to a divalent $Sr^{2+}$ ion.

A study of the magnetic spin texture of the ground state of the system
for one hole was previously discussed by one of us \cite {1skyrmion}, within
the framework of the $t - J$ model. There, the hole was only free to move
around one square plaquette, and it was found that the distorted AFM spins'
state could
be described, in a semi-classical sense \cite {shraimansiggia}, via a
non--coplanar spin state
with a rotational twist generated by the motional current of the hole
circulating around the plaquette. Since the hole can circulate in either the
clockwise or counter--clockwise directions, the ground state is two--fold
degenerate. The combination of the above results allows one to infer that the
spin texture is
analogous to a singly charged skyrmion found in classical chiral field theory
\cite {Belavin}. A numerical study \cite {bhattrabe} found similar results in
that
the symmetry of the ground state in their work is identical to that
of the skyrmion.

In order to model many such partially localised carriers an effective
Hamiltonian based on only the spin degrees of freedom was recently
proposed by two of us \cite {nskyrmions}. Again, based
on the semiclassical representation of the spins, a Hamiltonian
thought to mimick the effect of a hole confined to a single plaquette
was used:
\begin{eqnarray}
       H&=& J\sum_{\langle ij \rangle}  {\bf S}_{i}\cdot {\bf S}_{j}
            - \frac{D}{S^{4}}
        \left\{ ({\bf S}_{1}\cdot {\bf S}_{2}\times {\bf S}_{3})^{2} \right.
            \nonumber \\
        & & \left. +({\bf S}_{2}\cdot {\bf S}_{3}\times {\bf S}_{4})^{2}
        +({\bf S}_{3}\cdot {\bf S}_{4}\times {\bf S}_{1})^{2}
        +({\bf S}_{4}\cdot {\bf S}_{1}\times {\bf S}_{2})^{2}  \right\} .
\label{hamiltonian}
\end{eqnarray}
The first term is the usual Heisenberg interaction representing the
AFM background.  $J$ is the exchange coupling and $\langle ij \rangle$ denotes
nearest
neighbours in the plane; all near--neighbour pairs are summed over.
The second term is introduced to generate a spin distortion that is
the same as that produced by a hole circulating around the plaquette,
the corners of the plaquette being denoted by $1,2,3$ and $4$.
It was motivated by the work of Wen, Wilczek, and Zee \cite {wwz}
on chiral spin liquids, and we shall refer to this term as the
chiral--plaquette interaction.
The prefactor of the interaction, $D/S^{4}$, includes the $1/S^4$
factor to ensure that the Heisenberg and
one--plaquette interactions both scale as $S^2$. In Ref. \cite {nskyrmions}
it was found that the interaction strength $D$ must exceed $\sim 2.2~J$
to be able to induce a singly--charged skyrmion ground state.

In order to model a plane doped with many $O$ holes, one simply sums the
chiral--plaquette interactions over all plaquettes having a $Sr^{2+}$
ion above them; we distribute these defects randomly in the plane,
and the resulting system is necessarily spatially inhomogeneous.
The results of Monte Carlo studies of the spin--spin correlation function
\cite {class} for large lattices at temperatures greater than
$J/2$ were consistent with the empirical relation
\begin{equation}
\xi^{-1} (x,T) = \xi^{-1} (x,0) + \xi^{-1} (0,T)
\label{keimer}
\end{equation}
for the spin correlation length $\xi (x,T)$
found experimentally by Keimer {\it et~al.} \cite {keimer}. Note that
in order for Eq. (2.2) to be obtained, it was necessary to use $D \sim 3~J$.
Further, in Ref. \cite {nskyrmions} we showed that a mean field model of this
effective Hamiltonian was also consistent with Raman scattering
experiments \cite {sugai}. Now, in this paper, we extend our study
of this effective Hamiltonian and demonstrate that the transverse
spin--freezing temperature $T_f$ may be quantitatively determined.

\section{Semiclassical Evaluation of $T_{\lowercase {f}}$:}
\label{sect:semiclassical}

Here we present a discussion that leads to the analytic prediction for
the dependence of the spin--freezing temperature on doping.
The theory is based on a calculation
of the interaction energy between two skyrmions, and is performed
utilizing the semiclassical approach of Shraiman and Siggia \cite
{shraimansiggia}.
Two of us \cite {goodingmailhot} have previously used a similar approach
to calculate the interaction energy between two spin polarons generated
in the frustrated bond model \cite {aharony,frenkel}, a calculation
that was also verified via exact diagonalization numerical work \cite
{goodingmailhot}.

To begin, we introduce the spin texture associated with one skyrmion, where
the spins are described by a continuum field which is chiral. The direction
of the staggered magnetization is denoted by a chiral vector field $\bf {\hat
\Omega}$
satisfying
\begin{equation}
{\bf \hat \Omega \cdot \hat \Omega }= 1
\label{chiral}
\end{equation}
at every point in the plane.
For a singly charged skyrmion (Q=$\pm$ 1) placed at the origin of the
coordinate system
in use, and for one choice of phase (see below), where the staggered
magnetization
is chosen to be in the positive $z$ direction at infinity,
the chiral field is found to be \cite {Belavin}
\begin{equation}
{\bf \hat \Omega} = {2 \lambda x \over \lambda^2 + r^2}~{\bf \hat x}
\pm {2 \lambda y \over \lambda^2 + r^2}~{\bf \hat y}
+ {r^2 - \lambda^2 \over r^2 + \lambda^2}~{\bf \hat z}
\label{skyrmion}
\end{equation}
where $r = \sqrt {x^2 + y^2}$ is the distance in the two--dimensional plane
from the centre of the skyrmion, and $\lambda$ is the size of the skyrmion.
That $\lambda$ is the size of the skyrmion is seen from the $z$ component,
viz. for $r < \lambda$ the direction of the staggered magnetization is
reversed;
formally, $\lambda$ sets the length scale in the continuum problem. Further,
upon changing the topological charge Q from 1 to - 1, the handedness of
the twisting
of the staggered magnetization is found to change. This is seen in the
$x$ and $y$ components: note that ${\bf \hat \Omega}_x
\pm i {\bf \hat \Omega}_y \propto x \pm i y$, and thus the twisting
of the staggered magnetization is described by a vector which circulates in
the counter--clockwise (Q = 1) or clockwise (Q = -- 1) directions. Lastly, note
that these two components of the spin field represent the transverse degrees
of freedom and in this paper we shall suggest that it is these components
that lock together, due to the mutual interactions between skyrmions
mediated by the Heisenberg AFM superexchange,
leading to the spin freezing observed experimentally. A schematic
representation of the spin texture for the Q = 1 skyrmion is shown in Fig. 2.

We place one skyrmion at $\vec r = 0$, and then a second at $\vec r =
(\ell,0)$.
A charge of Q = 1 is chosen for the first, and the charge of the second will
be determined by minimizing the interaction energy. The latter may be obtained
from extrapolating the semiclassical theory \cite {shraimansiggia} to
describe this system. There, in the classical approximation, Shraiman
and Siggia derived that the interaction between a
dipolar spin distortion ${\bf P}_a$, such as that which would be produced by
a ferromagnetic bond in an AFM background (here $a = x$ or $y$
labels a direction
in the plane, such as the direction of the ferromagnetic bond),
and the surrounding spins is given by
\begin{equation}
E_{classical} = - g_2 \sum_{a =x,y} {\bf P}_a \cdot
\Big( {\bf \hat \Omega} \times \partial_a {\bf \hat \Omega} \Big) .
\label{Eclass}
\end{equation}
The interaction strength $g_2$ is a phenomenological
constant \cite {shraimansiggia}
which in the $t \gg J$ limit is of $O(J)$.
The ground state for the spin texture is then found by minimizing the
sum of this energy and the usual Heisenberg AFM interactions between spins,
the latter being most conveniently represented by the nonlinear sigma model.
We cannot associate a dipole moment with the spin texture of a skyrmion,
although
it can be represented as a superposition of dipolar backflow spin distortions
induced by mobile holes subsequently localised onto a single plaquette \cite
{sslong}.
A more direct way to represent the interaction of the background spin texture
with the
distortion induced by the skyrmion is to replace
\begin{equation}
{\bf P}_a \rightarrow {\bf \hat \Omega} \times \partial_a {\bf \hat \Omega}
\label{Pskyrmion}
\end{equation}
where the staggered magnetization in this equation is that of the skyrmion.
This
is the analogue of the definition of the semiclassical dipole moment given by
Shraiman and Siggia \cite {spiral} for mobile holes.
Then, we posit that one can represent the interaction between two skyrmions as
\begin{equation}
E_{int} = - g_2 \sum_{a =x,y}
\Big( {\bf \hat \Omega}^{(1)} \times \partial_a {\bf \hat \Omega}^{(1)} \Big)
\cdot
\Big( {\bf \hat \Omega}^{(2)} \times \partial_a {\bf \hat \Omega}^{(2)} \Big)
\label{Eint}
\end{equation}
where the superscripts 1 and 2 denote each of the two skyrmions. (Here we shall
assume that each of the spin fields describing a skyrmion are those
produced by systems with only one defect. Thus, no back--reaction effects
are included - these will be considered in the next section.)
As an estimate of this energy we evaluate $E_{int}$ at either of the points
in space where either of the two skyrmions are located.
In the limit of large $\ell$ we obtain
\begin{equation}
E_{int} = - 8 J / \ell^2, \quad \quad \ell \rightarrow \infty
\label{denergy}
\end{equation}
{\it when} the second skyrmion is chosen to be a Q = -- 1 spin distortion,
as this leads to the minimum energy. We note that the second skyrmion
can be placed {\it anywhere} in the lattice with the same result (viz. Eq.
(3.6))
being obtained as long as the second skyrmion has Q = -- 1. However, an
arbitrary phase of the skyrmion must be introduced (one that is not
included in Eq. (3.2) - see, e.g., Ref. \cite {Belavin} for such
a spin field), and then fixed via minimizing $E_{int}$ with respect
to this angle.  (The interaction energy between two skyrmions with the
same topological charge scales as $1 / \ell^4$ in the large $\ell$ limit.
As we will see below, this would lead to $T_f \propto x^2$, which
indeed is not observed experimentally.)

For holes randomly placed on a two--dimensional lattice one expects
that the doping $x$ would be given by $x = 1 / \ell^2$. This is
consistent with low temperature neutron scattering studies
of Keimer {\it et~al.} \cite {keimer} where the correlation
length $\xi (x,T)$ was found to be behave approximately as $\xi (x,0) \sim 1 /
\sqrt {x}$.
Thus, expressing the two--skyrmion interaction energy as a function
of doping, we finally have
\begin{equation}
E_{int} = - 8 J x .
\label{T_f.1}
\end{equation}

Now, we suggest that one may associate the spin--freezing temperature $T_f$
with this interaction energy; to be specific, $k_BT_f \propto | E_{int} |$.
This is consistent with the notion that there is only short--ranged ordering of
the
transverse spins (similar to the results of Ref. \cite {thomson}) and the
temperature at which the freezing occurs is such that
at higher temperatures thermal fluctuations would destabilize a spin texture
having these transverse degrees of freedom fixed.
Of course, the energy associated with when the transverse degrees of freedom
would become frozen into their ground state involves many skyrmions, and
thus we can only consider our treatment to be valid in the low doping limit.
Noting that the spins are $S = 1/2$ quantum spins, and thus we must include
the renormalization factor $Z_c$ accounting for quantum fluctuations,
we finally have that
\begin{equation}
k_B T_f \propto   8 J Z_c x
\label{T_f.2}
\end{equation}
where $Z_c \sim 1.2$  \cite {barnes}.
This relationship demonstrates that if the transverse degrees of freedom which
freeze are associated
with the many skyrmion spin texture produced by randomly distributed defects, a
linear dependence
of $T_f$ on doping is expected. This is entirely consistent with the
experiments
of Chou $et~al.$ \cite {chou}.

\section{Numerical Evaluation of $T_{\lowercase {f}}$:}
\label{sect:numerical }

We have just seen that the freezing of the transverse degrees
of freedom of two skyrmions separated by a distance $1 / \sqrt {x}$
would lead to a spin--freezing temperature linear in doping $x$.
In this section we present a numerical procedure to obtain the quantitative
dependence of $T_f$
on the doping again in the low temperature and lightly doped regime. This
work is required since the relationship found in the previous section,
viz. Eq. (3.8), did not allow for (i) the back--reaction, and subsequent
alteration, of a skyrmion's core, and thus must be an overestimate
of the interaction energy, and (ii) relies on an assumption for
$g_2$, the interaction strength. Here we wish to eliminate these
limitations and determine a very accurate numerical relationship
for $T_f$ on $x$.

To establish a relationship between $E_{int}$ and $x$ we will consider a
{\it lattice} of classical spins with two skyrmions placed somewhere
in the lattice a distance $\ell$ lattice spacings apart.
We calculate the interaction energy for such configurations
by evaluating the ground state using Eq. (2.1) as the Hamiltonian
(we have done the calculations for $S = 1$, and properly convert to
$S = {1\over 2}$ below).
In Ref. \cite {nskyrmions} it was found that the constant $D$ was
required to be $2.5< D/J < 3.3$ in order to explain both
the Raman scattering data \cite {sugai} and the neutron scattering
experiments consistent with Eq. (2.2). Here we perform the evaluation of
$E_{int}$ for $D/J = 2.5, 3,$ and $3.3$ in Eq. (2.1) to see how
$E_{int}$ varies with $D/J$.

The method employed in the evaluation of $E_{int}$ is straightforward and
involves studying $L \times L$ lattices with open boundary conditions
with two chiral--plaquette interactions included on two plaquettes, and
then extrapolating to $L \rightarrow \infty$;
for concreteness, we explain the method by considering Fig. 3. In this figure
we have shown
a $6 \times 6$ lattice and have labeled two plaquettes S$_1$
and S$_2$, where the separation between the centres of these plaquettes
correspond to $\ell = 2$. Firstly, we find the single skyrmion energy for
finite $L$
by eliminating the chiral--plaquette
interaction on S$_2$, and then find the ground state spin configuration
and associated energy (with open boundary conditions) - denote this
energy by $E_1 (L, \ell)$, where the inclusion of $\ell$ in $E_1$
indicates that the skyrmion is off centre on the lattice.
Then, the chiral--plaquette interaction on S$_2$ is reintroduced
and the two--skyrmion energy may be found
once the two--skyrmion ground state energy, denoted by $E_2 (L, \ell)$,
is known (again evaluated for open boundary conditions). All zero--force
spin configurations were evaluated using a conjugate--force method
similar to that employed in Ref. \cite {goodingmailhot}.

Consistent with the above semiclassical work, we always find that
the ground state for two defects corresponds to opposite
topological charges on the plaquettes.  Then, in analogy to a two--particle
binding energy, one defines
\begin{equation}
        E_{int}(L,\ell) = E_2(L,\ell) - 2 ( E_1 (L,\ell) - E_0 (L) ) - E_0 (L).
\label{binding}
\end{equation}
where $E_0(L)$ is the energy of a pure Heisenberg AFM, with open boundary
conditions, for an $L~\times~L$ lattice.
Finally, the large $L$ limit is taken to evaluate the interaction energy.
An example of this extrapolation is shown in Fig. 4 for $D/J = 3$ and
$\ell = 8 \sqrt {2}$, the latter corresponding to $x = .0078$.

In Fig. 5 we have plotted $| E_{int} |/J$ vs $x$, utilizing $\ell = 1 / \sqrt
{x}$,
for all three ratios of $D/J$. We obtain $5.14x \pm 0.08,~5.00x \pm .08,$ and
$4.74x \pm 0.08$,
for $D/J = 3.3,~3.0$ and $2.5$, respectively. The linear relation found in Eq.
(3.8)
is also found  in the numerical work, and thus validates the hydrodynamic
description of the spin degrees of freedom employed in the previous section.
Further,
it is clear that Eq. (3.7) represents the $D \rightarrow \infty$ limit (viz.
no alteration of the skyrmions' cores), and thus the energies found in this
section are appropriately less than that of Eq. (3.7). It is encouraging that
these energies do not change that much with $D/J$.

To determine the quantitative behaviour of $T_f(x)$ we must (i) transform
the above energies to represent those of $S = {1\over 2}$ quantum spins,
and (ii) determine the ratio of $| E_{int} |$ to $k_BT_f$. The first
point may be incorporated by multiplying the interaction energies by
$S(S+1)~Z_c$. The second is a nontrivial exercise, and to resolve it one
can appeal to experiment. Chou $et~al.$ \cite {chou} investigated
how best to describe the low--temperature maxima they observed
in the NQR rate, and found that if one attempted to describe
the maxima by a non--equilibrium freezing of the spins, with a correlation
time $\tau$ associated with an Arrhenius law, viz.
\begin{equation}
\tau \propto \exp (E / k_BT) ,
\end{equation}
then the relaxation rate peaked at $T_f$ with a related exponent
\begin{equation}
(E / k_BT_f) = 8.9 \pm 0.5
\end{equation}
(see Fig. 3 of Ref. \cite {chou}).
Clearly, the energy $E$ in the exponent in Eq. (4.3) must be associated
with an interaction energy, and thus we incorporate $| E_{int} |$ into this
equation. A theoretical explanation for this relationship follows from
application of the equipartition theorem. Apart from continuous symmetries,
there are three differing low--energy states of the single defect problem:
Q = $\pm~1$, as well as an excitation with zero topological charge
(see Ref. \cite {1skyrmion} for a discussion of this important
low--energy excitation). Thus, there are
$3^2$ {\it static} degrees of freedom for the two defect system. However,
since the spins are quantum spins, one should include a kinetic or {\it
dynamical}
component for the quantum fluctuations associated
with each of these degrees of freedom. Then, counting ${1\over 2} k_BT$
for each such degree of freedom one expects
\begin{equation}
E_{int} = 2~3^2~{1\over 2} k_BT_f .
\end{equation}
This is consistent with the experimental determination of Chou $et~al.$
given in Eq. (4.3).

Then, we immediately have estimates for the spin--freezing temperature:
\begin{equation}
T_f \sim (805~K)~x, {\rm ~for~} D = 3.3
\end{equation}
\begin{equation}
T_f \sim (784~K)~x, {\rm ~for~} D = 3.0
\end{equation}
\begin{equation}
T_f \sim (743~K)~x, {\rm ~for~} D = 2.5
\end{equation}
where we have used $J \sim 1550~K$ for the exchange constant. We anticipate
error bars of at least 5\% on the above values, the major uncertainties
coming from $Z_c$, and the exchange constant $J$.

The experimental result, viz.
$T_f \sim (815~K)~x$, shows that for $D$ being close to 3 this two--skyrmion
interaction model provides a very credible estimate of the spin--freezing
temperature. We stress that the temperature and doping dependence of
the spin correlation length was found to agree with Ref. \cite {keimer},
and the Raman scattering spectra of Ref. \cite {sugai}, {\it when}
the interaction strength $D \sim 3$ was used. It is gratifying that
quantitative
aspects of all three experiments seem to be consistent with the spin
texture found when the low--temperature doped planes are described
by Eq. (2.1).

\section{Discussion:}
\label{sec:discussion}

We have considered a simple model of the ground state spin texture produced
by a $Sr^{2+}$ defect and a single hole localised in its region \cite
{1skyrmion},
and then utilized a model that allows for the description of many such
defects \cite {nskyrmions}, viz. Eq. (2.1), where the chiral--plaquette
interaction is summed over all plaquettes having a divalent $Sr$
above them. We have used a semiclassical theory, and a numerical simulation
technique, to determine the interaction energy between two such skyrmion
states. We find that the ground state of such a system corresponds
to two skyrmions having opposite topological charges, and that then the
interaction energy, analogous to a two--particle binding energy,
is proportional to $1 / \ell^2$, as in Eq. (3.6), where $\ell$ is the
separation between defects. Then, we associate the doping $x$ with
this average distance via $\ell = 1 / \sqrt {x}$, consistent with
the low--temperature spin correlation length determined by neutron
scattering studies. The interaction energy is then found to scale
linearly with doping, and if we associate the spin--freezing temperature
with this interaction energy we can reproduce the same linear dependence that
is observed experimentally.
We have quantified this relationship by using the experimental value for
the Arrhenius exponent given in Eq. (4.3). We can also present theoretical
arguments that lead to values for the ratio of $E$ to $k_BT_f$ that give
the same numerical value as in Eq. (4.3) (e.g., a simple equipartition
theorem argument also gives a ratio of 9).

Alternatively, the frustrated bond model of Aharony $et~al.$ \cite {aharony}
predicts
an interaction energy which scales linearly as $x$ (see Ref. \cite
{goodingmailhot}),
but the spin texture of the ground state depends on the location of the
frustrated bonds
in the plane (e.g., see Ref. \cite {goodingmailhot}). Also,
the ratio of Eq. (4.4) would be 4 for the frustrated bond model,
and not $\sim$ 9, as found experimentally.
However, it is not clear whether or not the experimental results of
Chou $et~al.$ \cite {chou} depend on probe frequency, and thus it is
not clear if
this particular kind of experiment could distinguish between these two models
of the spin texture of the weakly doped planes. Since the important
physics of both these models is that defects produce long--ranged
spin distortions, and there is some evidence that such distortions
are also produced by mobile holes at moderate doping levels \cite
{nskyrmions,vos},
it is encouraging that both models rely on spin deviations coupling
the perturbing effects of the holes to the long--wavelength spin waves,
thus further justifying our hydrodynamic approach.

We have showed that for our model, with the interaction strength $D/J \sim 3$
in Eq. (2.1), the dependence of $T_f$ on $x$ is reproduced quite accurately.
Further, since our model with this same ratio of $D/J$ also reproduces the
neutron
scattering measurements of the spin correlation length \cite {keimer,shirane}
and the
Raman scattering spectra \cite {sugai}, it seems likely that these novel
topological excitations
of the $S = {1\over 2}$ quantum AFM do indeed play a role in forming
the spin textures of the low--temperature, inhomogeneously doped $CuO_2$
planes.

We wish to thank Fangcheng Chou, Ferdinand Borsa, and David Johnston for
valuable
discussions. This work was supported by the NSERC of Canada.

\begin{figure}
\noindent
1. Schematic phase diagram of the weakly doped system, showing
that at temperatures below $T_f(x)$ a coexisting quantum AFM with
long--ranged order and a spin--glass phase is observed.
\label{phasediagram}
\end{figure}

\begin{figure}
\noindent
2. Spin texture of a skyrmion state, such as that given in Eq. (3.2),
shown on a small $6~\times~6$ cluster
(here, Q=1, and $\lambda = .5$). The spin
state shown in the figure is found by taking Eq (3.2) and rotating
all spins (in spin space) by $\pi / 2$ along the $+z$ direction.
The spins at infinity are chosen to all point
out of the page, and we have only displayed the projection of the
spins onto the plane. Also, for ease of presentation, all spins on
B sublattice sites of the background
N\'eel state have been inverted ($\vec S_i \rightarrow - \vec S_i$),
making the $D = 0$ state be a completely polarized ferromagnetic state.
The solid dots denote the lattice sites.
\label{singleskyrmion}
\end{figure}

\begin{figure}
\noindent
3. Schematic diagram of a small cluster denoting two $Sr$ defects
above plaquettes $S_1$ and $S_2$.
\label{twoskyrms}
\end{figure}

\begin{figure}
\noindent
4. Interaction energy defined in Eq. (4.1), in units of $J$, for
$D = 3$ and a separation of defects of $8\sqrt 2$ (viz., eight diagonal
spacings), as a function of system size $L$ for $L$ = 20, 24, and 28,
showing the extrapolation to the bulk limit.
\label{eintfig1}
\end{figure}

\begin{figure}
\noindent
5. The magnitude of the interaction energy $| E_{int} |$, in units
of $J$, as a function of doping $x$ for differing ratios of $D / J$.
For each $D / J$, the data extrapolates linearly back to the origin.
\label{eintfig2}
\end{figure}


\begin{references}
\bibitem{nishihara} H. Nishihara, H. Kasuoka, T. Schimizu, T. Tsuda,
T. Imai, S. Sasaki, S. Kanbe, K. Kishio, K. Kitazawa, and K. Fueki,
J. Phys. Soc. Jpn. {\bf 56},
4559 (1987).
\bibitem{borsa} F. Borsa, M. Corti, T. Rega, and A. Rigamonti,
Nuovo Cimento D {\bf 11}, 1785 (1989).
\bibitem{chou} F.C. Chou, F. Borsa, J.H. Cho, D.C. Johnston,
A. Lasciarlfari, D.R. Torgeson, and J. Ziolo, Phys. Rev. Lett. {\bf 71}, 2323
(1993).
\bibitem{chen} M.C. Chen and C.P. Slichter, Phys. Rev. B {\bf 27}, 278 (1983).
\bibitem{cho} J.H. Cho, F.C Chou, and D.C. Johnston, Phys. Rev. Lett. {\bf 70},
222 (1993).
\bibitem{chouthesis} F.C. Chou, Ph.D. Thesis, Iowa State University
(unpublished).
\bibitem{watanabe} I. Watanabe, K. Kumagai, Y. Nakamura, and
H. Nakajima, J. Phys. Soc. Jpn. {\bf 59}, 1932 (1990).
\bibitem{rettori} C. Rettori, D. Rao, S.B. Oseroff, G. Amoretti, Z. Fisk,
S. Cheong, D. Vier, R.D. Zysler, and J.E. Schirber, Phys. Rev. B {\bf 47}, 8156
(1993).
\bibitem{ryan} D.H. Ryan, J.O. Str\"om--Olsen, W.B. Muir, J.M. Cadogan,
and, J.M.D. Coey, Phys. Rev. B {\bf 40}, 11208 (1990).
\bibitem{thomson} J.R. Thomson, H. Guo, D.H. Ryan, M.J. Zuckerman,
and M. Grant, Phys. Rev. B {\bf 45}, 3129 (1992);
J.R. Thomson, Ph.D. Thesis, McGill University (1992).
\bibitem{1skyrmion} R.J. Gooding, Phys. Rev. Lett. {\bf 66}, 2266 (1991).
\bibitem{nskyrmions} R.J. Gooding, and A. Mailhot, Phys. Rev. B, Phys. Rev. B
{\bf 48}, 6132 (1993).
\bibitem{resistivity} C.Y Chen, R.J. Birgeneau, M.A. Kastner, N.W. Preyer,
and T. Thio, Phys. Rev. B {\bf 43}, 392 (1991).
\bibitem{shraimansiggia} B.I. Shraiman, and E.D. Siggia, Phys. Rev. Lett. {\bf
61},
467 (1988).
\bibitem{Belavin} A.A. Belavin, and A.M. Polyakov, JETP Lett. {\bf 22}, 245
(1975).
\bibitem{bhattrabe} K.M. Rabe, and R.N Bhatt, J. Appl. Phys. {\bf 69}, 4508
(1991).
\bibitem{wwz} X.G. Wen, F. Wilczek, and A. Zee, Phys. Rev. B {\bf 39}, 11413
(1991).
\bibitem{class} Again, this is done for classical spins, and assumes that for
the static properties one is still in the renormalized classical regime for
weakly doped planes.
\bibitem{keimer} B. Keimer, N. Belk, R.J. Birgeneau, A. Cassanho, C.Y. Chen,
M. Greven, M.A. Kastner, A. Aharony, Y. Endoh, R.W. Erwin, and G. Shirane,
Phys. Rev. B {\bf 46}, 14034 (1992).
\bibitem{sugai} S. Sugai, S.I. Shamoto, and M. Sato, Phys. Rev. B. {\bf 38},
6436 (1988).
\bibitem{goodingmailhot} R.J. Gooding, and A. Mailhot, Phys. Rev. B {\bf 44},
11852 (1991).
\bibitem{aharony} A. Aharony, R.J. Birgeneau, A. Coniglio, M.A. Kastner,
and H.E. Stanley, Phys. Rev. Lett. {\bf 60}, 1330 (1988).
\bibitem{frenkel} D.M. Frenkel, R.J. Gooding, B.I. Shraiman, and E.D. Siggia,
Phys. Rev. B {\bf 41}, 350 (1990).
\bibitem{sslong} B.I. Shraiman, and E.D. Siggia, Phys. Rev. B {\bf 42}, 2485
(1990).
\bibitem{spiral} B.I. Shraiman, and E.D. Siggia, Phys. Rev. Lett. {\bf 62},
1564 (1989).
\bibitem{barnes} T. Barnes, Int'l J. Mod. Phys. C {\bf 2}, 659 (1991).
\bibitem{vos} R.J. Gooding, K.J.E. Vos, and P.W. Leung, to be published, Phys
Rev. B, Feb. 94.
\bibitem{shirane} G. Shirane, R.J. Birgeneau, Y. Endoh, and M.A. Kastner,
Physica B (to be published).
\end{references}
\end{document}